\documentclass[11pt]{article}
\usepackage[dvipdfm]{graphicx}
\usepackage{amsmath,amssymb} 

\begin{document}

\baselineskip=11.0pt

\centerline{Hardon-quark hybrid stars constructed by}

\centerline{the nonlinear $\sigma$-$\omega$-$\rho$ mean-field model and MIT-bag model}

\vspace{0.5cm}

\centerline{Schun~T.~Uechi\footnote{E-mail: suechi@rcnp.osaka-u.ac.jp} and Hiroshi Uechi\footnote{E-mail: uechi@ogu.ac.jp}}
\vspace{0.5cm}

\centerline{$^1$Research Center for Nuclear Physics ($RCNP$), Osaka University, Ibaraki, Osaka 567-0047 \quad Japan}

\centerline{$^2$Department of Distributions and Communication Sciences, Osaka Gakuin University}
\centerline{2-36-1  Kishibe-minami,  Suita,  Osaka  564-8511 \quad Japan}

\vspace{1.0cm}

\makeatletter
\renewcommand{\theequation}{%
\thesection.\arabic{equation}}
\@addtoreset{equation}{section}
\makeatother

\newcommand{\bkappa}{\mbox{\boldmath $\kappa$}}
\newcommand{\btau}{\mbox{\boldmath $\tau$}}
\newcommand\kfermi{k_{\scriptscriptstyle F}}
\newcommand\kfermiB{k_{\scriptscriptstyle B}}
\newcommand\Mstar{M^{\ast}}
\newcommand\kstar{k^{\ast}}
\newcommand\rhoB{\rho_{\scriptscriptstyle B}}
\newcommand\szero{\scriptscriptstyle (0)}
\newcommand{\bftau}{\mbox{\boldmath $\tau$}}
\newcommand {\bR} {\mbox{\boldmath $R$}}
\newcommand {\bL} {\mbox{\boldmath $L$}}
\newcommand {\bV} {\mbox{\boldmath $V$}}

\centerline{\large{Abstract}}

Density-dependent relations among saturation properties of symmetric nuclear matter and hyperonic matter, properties of
hadron-(strange) quark hybrid
stars are discussed by applying the conserving nonlinear $\sigma$-$\omega$-$\rho$ hadronic mean-field theory.  Nonlinear interactions
that will be renormalized as effective coupling constants, effective masses and sources of equations of motion are constructed
self-consistently by maintaining thermodynamic consistency to the mean-field approximation.  The coupling constants expected from
the hadronic mean-field model and SU(6) quark model for the vector coupling constants
are compared; the coupling constants exhibit different density-dependent results for effective masses and binding energies of hyperons,
properties of hadron and hadron-quark stars.  The nonlinear $\sigma$-$\omega$-$\rho$ hadronic mean-field approximation with or
without vacuum fluctuation corrections and strange quark matter defined by MIT-bag model are employed to examine properties of
hadron-(strange) quark hybrid stars.  We have found that hadron-(strange) quark hybrid stars become more stable in high density
compared to pure hadronic and strange quark stars.

PACS numbers: 21.65.+f, 24.10.Cn, 24.10.Jv, 26.60.+C


\pagestyle{plain}
\setcounter{page}{1}

\section{Introduction}

The symmetric nuclear matter is a self-bound matter whose binding energy exhibits a characteristic concave curve at saturation density,
where pressure of nuclear matter vanishes ($p=0$).  It has been known as a constraint to examine self-consistency to nuclear
many-body approximations~$\cite{DAY}-\cite{KAB}$.   The self-consistency has been fundamental for many-body approximations
in terms of Landau's Fermi-liquid theory~$\cite{LAN}-\cite{NOZ}$, Kadanoff-Baym's theory of conserving
approximations~$\cite{BAY}-\cite{BON}$ and the density functional theory~$\cite{WKS}-\cite{UEC1}$.  The relativistic,
field-theoretical, linear $\sigma$-$\omega$ mean-field approximations that maintain conditions of conserving approximations
have been applied to finite nuclei, nuclear matter and neutron stars~$\cite{WAL,SER}$.  These nuclear mean-field approximations
reveal an important relation among self-consistent single particle energy, energy density and particle density, which is denoted as
{\it thermodynamic consistency}~$\cite{UEC3,FUS}$.   The conserving nonlinear
$\sigma$-$\omega$-$\rho$ mean-field approximation which maintains thermodynamic consistency has been applied
to investigate density-dependent correlations among properties of nuclear, neutron and hyperonic
matter~$\cite{UEC2}-\cite{HUS}$.  Thermodynamic consistency is important to derive consistent results for high energy and
high density phenomena, such as heavy-ion collision, hypernuclei-formation experiments~$\cite{ASB}$ and astrophysical
phenomena~$\cite{WEI}-\cite{GLE}$.

The nonlinear $\sigma$-$\omega$-$\rho$ mean-field approximation has several nonlinear coefficients whose values are determined
by reproducing binding energy, symmetry energy and simultaneously searching the minimum value of incompressibility at saturation
density, as well as reproducing a maximum mass of neutron stars at high density.  The binding energy
at saturation is taken as $-15.75$ MeV at $\kfermi = 1.30$ fm$^{-1}$ ($\rho_0 = 0.148$ fm$^{-3}$) with
the symmetry energy, $a_4 = 30.0$ MeV in the current calculation~$\cite{WAL}$.  The values of nonlinear coefficients
which will produce effective masses of hadrons $\Mstar_N/M_N \sim 0.70$, $m_{\sigma}^{\ast}/m_{\sigma} \sim 1.02$,
$m_{\omega}^{\ast}/m_{\omega} \sim 1.01$ and incompressibility, $K \sim 320$ MeV, with symmetry energy $a_4 \sim 30$ MeV,
the maximum mass of neutron stars in $\beta$-equilibrium matter~$\cite{GLE}$, $M_{max}(n,p,e) \sim 2.50$ $M_{\odot}$
(the solar mass: $M_{\odot} \sim 1.989 \times 10^{30}\ {\rm kg}$), are found reasonable and admissible in the nonlinear
$\sigma$-$\omega$-$\rho$ Hartree approximation~$\cite{SCH,HUS}$.

The maximum masses of neutron stars are expected to be below
$2.5\ M_{\odot}$~$\cite{AKM}$, or $2.1 \pm 0.2 M_{\odot}$~$\cite{LAT}$, which depends on the hadronic equation of state (EOS) for
isospin asymmetric, hyperonic matter in the density range, $2\rho_0 \lesssim \rhoB \lesssim 5\rho_0$.   The hadronic
liquid-gas phase transition at the surface of neutron
stars and determination of the radius of smeared, gas-phase surface are not directly relevant to properties of hadronic matter
and maximum masses of neutron stars.   The nonlinear interactions exhibit significant density-dependent effects on
incompressibility, $K$, and symmetry energy, $a_4$, in high densities; these Fermi liquid properties monotonically
increase about saturation density, but they are piecewise continuously softened at an hyperon onset density.  These phenomena of
piecewise continuous change of Fermi-liquid properties will be important for the analysis of Landau parameters, heavy-ion collision,
high energy and high density experiments.  At a hyperon onset density from ($n, p, e$) to ($n, p, H, e$), the EOS suddenly
becomes softer.  This is because the nucleon Fermi energy, $E_N (\kfermi)$ in the phase ($n, p, e$), will be redistributed
to the hyperon Fermi energy, $E_H (\kfermi)$ in the phase ($n, p, H, e$); consequently, the Fermi energies become relatively
small in the phase ($n, p, H, e$) compared with those of ($n, p, e$).  The redistribution and slow increase of
Fermi energies appear whenever hyperons are generated, resulting in a softer equation of state and discontinuous changes of $K$ and
$a_4$.  This is numerically checked as discrete changes of physical quantities, such as effective masses of
hadrons, incompressibility, symmetry energy and energy density~$\cite{SCH,HUS}$.

Since the hyperon-onset will confine Fermi-energies of baryons as explained above, single and double hyperon generations
exhibit different density-dependent phenomena.  For example, the $\Lambda$-hyperon onset density such as
in ($n, p, e$)-($n, p, \Lambda, e$) is $\rho_{\Lambda}/\rho_0 \sim 2.2$; however, it gives $\rho_{\Lambda}/\rho_0 \sim 4.2$
when $\Lambda$ is generated in ($n, p, e$)-($n, p, \Sigma^{-}, e$)-($n, p, \Sigma^{-}, \Lambda$).  The same phenomena
are observed with other hyperons, and generally the onset-density of a hyperon is pushed up to
a higher density, which is denoted as the {\it push-up phenomena} of hyperon onset-densities in many-fold hyperon 
generations~$\cite{SCH}$.  Because of the push-up phenomena of the hyperon onset density, we have found that hyperon generations
are suppressed in high densities and hyperons relevant to determine the maximum mass of neutron stars are
$\Sigma^{-}$ and $\Lambda$.  The similar results are discussed in nonrelativistic Brueckner-Hartree-Fock~$\cite{IVI,BBS}$ and
quark matter calculations~$\cite{TMA}$.  The hyperon onset densities are related to hyperon single particle energies by way of
phase equilibrium conditions.   Hence, the Hugenholtz-Van Hove theorem and thermodynamic consistency at saturation of hyperon
binding energy are essential to define self-consistent approximation of hyperonic matter.  The single particle energies of
hyperons are also important to study $K$ and $a_4$ for magic nuclei~$\cite{HUE3}$ and hypernuclei~$\cite{AGA}-\cite{HSH}$.

Density-dependent effective masses and effective coupling constants, saturation properties for nucleons and hyperons are
discussed in the nonlinear $\sigma$-$\omega$-$\rho$ mean-field approximation~$\cite{SCH,HUS}$.  The results show;
(1) coupling constants of hyperons are related to those of nucleons by effective masses, effective coupling constants, binding energy
at onset-density of respective hyperons.  Hence, it shows that binding energies of symmetric nuclear matter and hyperon matter are
self-consistently related to each other.  (2) Self-consistency suppresses hyperon generations in high densities, denoted as
{\it the push-up phenomena} of hyperon onset densities.  The suppression of hyperon generations is also discussed in different
calculations~$\cite{IVI}-\cite{TMA}$.  (3) Coupling ratios of hyperons are expected to be
$g_{\sigma \scriptscriptstyle H}/g_{\sigma \scriptscriptstyle N},
g_{\omega \scriptscriptstyle H}/g_{\omega \scriptscriptstyle N} \gtrsim 1$, in order to be consistent with conditions of
thermodynamic consistency, empirical values of nuclear matter and neutron stars.  In the current calculations,
we have included vacuum fluctuation corrections (VFC) into the nonlinear $\sigma$-$\omega$-$\rho$ approximation and
examined properties of ($n, p, e$), ($n, p, H_1, e$), ($n, p, H_1, H_2, e$) hyperonic matter, constrained by transitions
to strange quark matter and properties of hadron-quark hybrid stars.   The hadron-quark phase transitions
are assumed to be a first order and computed by Maxwell construction~$\cite{MBE}$.

The conserving nonlinear $\sigma$-$\omega$-$\rho$ mean-field approximation~$\cite{UEC2}-\cite{HUS}$ and
quark-based effective nuclear models~$\cite{GGE}-\cite{AHT}$ have been applied to
finite nuclei, nuclear and isospin asymmetric, high-density matter.  We have compared hyperon coupling constants
required by the nonlinear $\sigma$-$\omega$-$\rho$ mean-field model with those required by SU(6) quark model for the vector coupling
constants~$\cite{JSI,GSY}$.   The hyperon coupling constants required by hadronic and SU(6) quark models exhibit
quite different results for effective masses, binding energies of hyperons~$\cite{SCH,HUS}$ and
properties of hadron-quark hybrid stars.  The coupling ratios required by SU(6) quark model cannot reproduce hyperon
saturation properties, which will be discussed in terms of conditions of thermodynamic
consistency in the sec.~6.  The analysis of discrepancies of predictions by hadronic and
quark-based models may provide us with insight for constructing self-consistent nuclear many-body
problems~$\cite{SER}-\cite{HUS}$.  The effective masses of hyperons depend on coupling
ratios ($g_{\sigma H}/g_{\sigma N} \equiv r^{\sigma}_{H N}$, $g_{\omega H}/g_{\omega N} \equiv r^{\omega}_{H N}$,
$g_{\rho H}/g_{\rho N} \equiv r^{\rho}_{H N}$) and decrease analogous to effective masses of nucleons in high
densities, which shows strong density-dependency of hadronic interactions.  However, the effective masses of hyperons
exhibit weak density-dependent results with $r^{\omega}_{H N} \leq 2/3$
required by SU(6) quark model for the vector coupling constants.  In general, hadrons exhibit strong
density-dependent interactions and correlations among properties of nuclear matter, hyperonic matter and neutron stars.  In the
hadronic mean-field approximation, $r^{\omega}_{H N} \gtrsim 1.0$ is preferred in order to be consistent with properties of
nuclear matter and the maximum mass of isospin asymmetric neutron stars, $M_{max}(n,p,e) = 2.00 \sim 2.50$ $M_{\odot}$.  The
value, $r^{\omega}_{H N} = 2/3$, results in different effective masses and binding energies of hyperons; the discrepancies
originate from density-dependent interactions of hadrons~$\cite{HUS}$.

The density-dependent many-body effects produced by the conserving nonlinear $\sigma$-$\omega$-$\rho$ mean-field approximation should be
compared to chiral hadronic mean-filed approximations~$\cite{FST2}-\cite{TKJ}$.  The effective quark-based
chiral lagrangian approach suggests that the appropriate in-medium scaling law, $m^{\ast}_{\sigma}/m_{\sigma} \approx m^{\ast}_N/m_N
\approx m^{\ast}_{\omega}/m_{\omega} \approx f^{\ast}_{\pi}/f_{\pi} \approx m^{\ast}_{\rho}/m_{\rho}$, be
expected~$\cite{GEB,RHO}$.  Although it is ambiguous whether $m^{\ast}_N/m_N \approx m^{\ast}_{\omega}/m_{\omega}$ means
that $m^{\ast}_{\omega}/m_{\omega} \lesssim 1.0$ or $m^{\ast}_{\omega}/m_{\omega} \gtrsim 1.0$ at saturation density,
they are certainly decreasing above saturation density, since $m^{\ast}_N/m_N$ is model-independently expected to be decreasing.  This is
significantly different from nonlinear mean-field approximations, since self-consistency of hadronic mean-field approximation
demands $m^{\ast}_{\sigma}/m_{\sigma} \approx m^{\ast}_{\omega}/m_{\omega} \gtrsim 1.0$~$\cite{UEC2,MCH}$ and
$m^{\ast}_{\rho}/m_{\rho} \approx m^{\ast}_{\pi}/m_{\pi} \lesssim 1.0$~$\cite{PSJ}$ in high densities.  In other words,
since effective masses and coupling constants are self-consistently related to each other, if some values decrease,
the others have to counterbalance the variations.  Therefore, all correlated effective masses and coupling constants must
decrease or increase simultaneously in order to compensate for variations among others.  At saturation density, effective masses of
mesons are density-dependent and their ratios are approximately equal:
$m^{\ast}_{\sigma}/m_{\sigma} \approx m^{\ast}_{\omega}/m_{\omega} \approx m^{\ast}_{\rho}/m_{\rho} \approx
m^{\ast}_{\pi}/m_{\pi} \approx 1.0$.  However, mesons exhibit different behavior above saturation densities as
explained.  The discrepancy has been discussed in terms of thermodynamic consistency in the hadronic mean-field
approximations~$\cite{UEC2}-\cite{HUS}$ and should be investigated in other
hadronic models to extract consistent model-independent results.

We have applied the nonlinear $\sigma$-$\omega$-$\rho$ mean-field approximation and MIT-bag model upon hadron-quark hybrid
stars~$\cite{GLE,BDS}$.  The numerical analysis exhibits new results that the hadron-(strange) quark hybrid stars are more stable
in high density than pure hadronic and strange quark stars.  It suggests a relation between bag constant and
QCD strong coupling constant, ($B$, $\alpha_c$)~$\cite{ACH,EFA}$, to the central density and maximum mass of
hadron-(strange) quark hybrid stars, (${\cal E}_c$, $M_{max}$).  The results obtained in the current calculations should be
examined from astronomical data whether or not hadron-quark stars are possible and the values of bag constant
and strong coupling constant, ($B$, $\alpha_c$), could be consistent with astronomical data for neutron stars.

Self-consistent relations among saturation properties of nucleons and hyperons are briefly reviewed in sec.~2.  Quantitative
numerical calculations for effective masses, onset densities and conditions of hyperon saturation are discussed in
the articles~$\cite{SCH,HUS}$.  The MIT-bag quark matter and vacuum fluctuation
correction to nonlinear mean-field approximation are explained in sec.~3.  Results of pure hadron and
hadron-strange quark hybrid stars are discussed in sec.~4; concluding remarks are in sec.~5.

\section{Self-consistent effective masses and coupling constants in the nonlinear $\sigma$-$\omega$-$\rho$ mean-field approximation}

The hadronic lagrangian with nonlinear $\sigma$-$\omega$-$\rho$ interactions which yields density-dependent effective
masses and coupling constants is given by~$\cite{UEC2}$,

\begin{eqnarray}
&& {\cal L} =\sum _{B}\bar {\psi }_{B}[\gamma _{\mu }(i\partial ^{\mu }-g_{\omega B}^{\ast }V^{\mu}
-\frac{g_{\rho B}^{\ast}}{2}\btau\cdot \bR ^{\mu})-(M_{B}-g^{\ast}_{\sigma B}\phi)]\psi_{B}\nonumber \\
&& \hspace{0.8cm}+\frac{1}{2}(\partial_{\mu }\phi \partial ^{\mu}\phi -m_{\sigma}^{2}\phi^{2})-\frac{g_{\sigma 3}}{3!}\phi^{3}
-\frac{g_{\sigma 4}}{4!}\phi^{4}\nonumber \\
&& \hspace{0.8cm}-\frac{1}{4}F_{\mu \nu }F^{\mu \nu}+\frac{1}{2}m_{\omega}^{2}V_{\mu }V^{\mu}
+\frac{g_{\omega 4}}{4!}(V_{\mu}V^{\mu})^{2}+\frac{g_{\sigma \omega}}{4}\phi^{2}V_{\mu}V^{\mu}\nonumber \\
&& \hspace{0.8cm}-\frac{1}{4}\bL_{\mu \nu}\cdot \bL^{\mu \nu }+\frac{1}{2}m_{\rho}^{2}\bR_{\mu}\cdot \bR^{\mu}
+\frac{g_{\rho 4}}{4!}(\bR_{\mu}\cdot \bR ^{\mu })^{2}+\frac{g_{\sigma \rho}}{4}\phi^{2}\bR_{\mu}\cdot \bR^{\mu}
+\frac{g_{\omega \rho}}{4}V_{\mu}V^{\mu}\bR_{\mu}\cdot \bR^{\mu} \nonumber \\
&& \hspace{0.8cm}+\sum_{l}{\bar {\psi}}_{l}(i\gamma_{\mu}\partial^{\mu}-m_{l})\psi_{l}+\delta{\cal L} \label{eqn:FLA}
\end{eqnarray}
where $\psi_B$ ($B = n, p, \Lambda, \Sigma, \cdots$) and $\psi_l$ ($l = e^{-}, \mu^{-}$) denote the field of baryons and leptons,
respectively.  The meson-fields operators are: $\phi $ for the $\sigma$-field, $V$ for the vector-isoscalar $\omega$-meson,
$V_{\mu} V^{\mu} = V_0^2 - \bV ^2$, $(\mu = 0, 1, 2, 3)$ and $R_{\mu}$ for $\rho$-meson.  The vector field strengths,
$F_{\mu \nu}$ and $\bL _{\mu \nu}$, are defined as, $F_{\mu \nu}=\partial_{\mu }V_{\nu}-\partial_{\nu}V_{\mu}$ and
$\bL_{\mu \nu}=\partial_{\mu}\bR_{\nu}-\partial_{\nu}\bR_{\mu}+g_{\rho} \bR_{\mu} \times \bR_{\nu}$.

The coupled nonlinear quantum-field lagrangian~($\ref{eqn:FLA}$) is interpreted as baryon quantum-field lagrangian in
mean fields of mesons~$\cite{WAL}$.  All dynamics generated by baryon fields are mediated by mean fields of mesons which will
be self-consistently defined in an approximation.  The effective coupling constants, $g_{i B}^{\ast}$ $(i=\sigma, \omega, \rho)$,
denote renormalized, density-dependent coupling constants defined by self-consistent mean-field of $\sigma$-meson.  The nonlinear
$\sigma$-$\omega$-$\rho$ mean-field model maintains the structure of Serot and Walecka's linear $\sigma$-$\omega$ mean-field
approximation~$\cite{WAL}$, Lorentz-invariance and renormalizability, the Hugenholtz-Van Hove theorem~$\cite{HUG}$,
conditions of conserving approximations~$\cite{BAY}-\cite{UEC1}$, the virial theorem~$\cite{UEC3,FUS}$ and
Landau's hypothesis of quasiparticles~$\cite{LAN}-\cite{NOZ}$.  As we proved in the ref.~$\cite{UEC2}$, nonlinear mean-field
approximations are equivalent to Hartree approximation when nonlinear interactions are properly renormalized, and consequently,
the concepts of {\it effective masses} and {\it effective coupling constants} are naturally generated by nonlinear mean-field
interactions.  Self-consistent relations among single particle energy, effective masses and coupling constants will restrict
empirical values of low-density nuclear matter and high-density hadronic matter.  The admissible values of effective coupling
constants and masses are confined in certain values due to strong density-dependent correlations among physical quantities
of nuclear matter and neutron stars~$\cite{UEC2}-\cite{HUS}$.

Meson-fields operators are replaced by mean-fields denoted by $\phi_0$, $V_0$, and $R_0$.  The equations of motion for baryons are given by,
\begin{equation}
\Bigl[ (i\gamma_{\mu} \partial^{\mu} - g_{\omega B}^{\ast} \gamma_0 V_0 - \frac{g_{\rho B}^{\ast}}{2} \gamma_0 \tau_3 R_0)
- (M_B - g_{\sigma B}^{\ast} \phi_0) \Bigr] \psi_B = 0 \label{eqn:EVXB} \ ,
\end{equation}
where $g_{\sigma B}^{\ast}$, $g_{\omega B}^{\ast}$ and $g_{\rho B}^{\ast}$ are {\it effective coupling constants} for $\sigma$, $\omega$ and
$\rho$ mesons.  One should notice that effective coupling constants cannot be simply introduced as experimental, external inputs to an employed
approximation, since density-dependent coupling constants will modify equations of motion for mesons.  We will assume that only
nucleon-meson coupling constants are density-dependent since we are interested in the density-dependent correlations among properties
of symmetric nuclear matter.  Nonlinear interactions are not assumed in the coupling constants for hyperons; effective masses of hyperons
are defined by $M^{\ast}_H = M_H - g_{\sigma H} \phi_0$.  The density-dependent nucleon-meson coupling constants that maintain
thermodynamic consistency are,
\begin{equation}
\begin{split}
g_{\sigma N}^{\ast} =&\ g_{\sigma N} + g_{\sigma \sigma N}\phi_0/2m_{\sigma} \ , \\
g_{\omega N}^{\ast} =&\ g_{\omega N} + g_{\sigma \omega N} \phi_0/m_{\sigma} \ , \\
g_{\rho N}^{\ast}/2 =&\ g_{\rho N}/2 + g_{\sigma \rho N} \phi_0/m_{\sigma} \ . \label{eqn:EFC}
\end{split}
\end{equation}
The equations of motion for mesons are given by,
\begin{eqnarray}
m_{\sigma}^2 \phi_0 + \frac{g_{\sigma 3}}{2!} \phi_0^2 + \frac{g_{\sigma 4}}{3!} \phi_0^3 - \frac{g_{\sigma \omega}}{2} V_0^2 \phi_0
- \frac{g_{\sigma \rho}}{2} R_0^2 \phi_0 - \frac{g_{\sigma \sigma N}}{2 m_{\sigma}} \rho_s \phi_0
=&& g_{\sigma N}^{\ast} \rho_s - \frac{g_{\sigma \omega N}}{m_{\sigma}} V_0 \rho_{\omega}
- \frac{g_{\sigma \rho N}}{m_{\sigma}} R_0 \rho_3 \ , \nonumber \\
m_{\omega}^2 V_0 + \frac{g_{\omega 4}}{3!} V_0^3 + \frac{g_{\sigma \omega}}{2} \phi_0^2 V_0 + \frac{g_{\omega \rho}}{2} R_0^2 V_0
=&& g_{\omega N}^{\ast} \rho_{\omega}  \ , \nonumber \\
m_{\rho}^2 R_0 + \frac{g_{\rho 4}}{3!} R_0^3 + \frac{g_{\sigma \rho}}{2} \phi_0^2 R_0 + \frac{g_{\omega \rho}}{2} V_0^2 R_0
=&& \frac{g_{\rho N}^{\ast}}{2} \rho_3 \label{eqn:EVXM} \ .
\end{eqnarray}
where $\rho_s = \sum_B \rho_{sB}$ is the total scalar source, $\rho_{\omega}$ the isoscalar density,
and $\rho_3 = (k_{F_p}^3 - k_{F_n}^3)/3\pi^2$, the isovector density.  The density-dependent coupling constants will modify
the equation of motion for $\sigma$-meson which acquires a mass term, $\frac{g_{\sigma \sigma N}}{2 m_{\sigma}} \rho_s \phi_0$,
and new source terms, $-g_{\sigma \omega N} V_0 \rho_{\omega}/m_{\sigma} - g_{\sigma \rho N} R_0 \rho_3/m_{\sigma}$ from
density-dependency of effective coupling constants.

The introduction of nonlinear $\sigma\sigma N$-vertex interaction leads to the effective mass of nucleon:
\begin{equation}
M^{\ast}_N = M_N - g_{\sigma N}^{\ast} \phi_0 = M_N - g_{\sigma N} \phi_0 - (g_{\sigma\sigma N}/2m_{\sigma}) \phi_0^2 \ ,
\label{eqn:PNM}
\end{equation}
and effective masses of nucleons and hyperons are related to each other as,
\begin{equation}
M_H - M^{\ast}_H = \frac{g_{\sigma H}}{g_{\sigma N}^{\ast}} (M_N - M_N^{\ast}) \ . \label{eqn:NHM}
\end{equation}
The total scalar source is obtained by the requirement of self-consistency,
\begin{equation}
\Sigma^s = \Sigma^s_N + \Sigma^s_H = - \frac{g_{\sigma N}^{\ast 2}}{m_{\sigma}^{\ast 2}} (\rho_{sN}^{\ast} + \rho_{sH}) \ ,
\label{eqn:TSS}
\end{equation}
where the scalar sources are respectively given by
\begin{equation}
\rho_{sB} = \frac{g_{\sigma B}/g_{\sigma N}^{\ast}}{\pi^2} \int^{k_{F_B}}_0 \! dq q^2 \frac{\Mstar_B}{E^{\ast}_B (q)} \ ,
\label{eqn:SSB}
\end{equation}
and $\rho_{sN}^{\ast}$ is the modified scalar density defined by $\displaystyle g_{\sigma N}^{\ast} \rho_{sN}^{\ast}
= g_{\sigma N}^{\ast} \rho_{sN} - g_{\sigma\omega N} V_0 \rhoB/m_{\sigma} - g_{\sigma\rho N} R_0 \rho_3/m_{\sigma}$;
$N$ is used to denote proton and neutron, $N=(p, n)$; hyperons are denoted as,
$H = \Lambda, \Sigma^{-}, \Sigma^{0}, \Sigma^{+}, \cdots$.  The $\omega$-meson and $\rho$-meson contributions to self-energies
are given by
\begin{equation}
\displaystyle \Sigma_{\omega}^{\mu} = - \frac{g_{\omega N}^{\ast 2}}{m_{\omega}^{\ast 2}} \rho_{\omega} \delta_{\mu, 0} \quad
{\rm and} \quad 
\Sigma^{\mu}_{\rho(^p_n)} = \mp \frac{g_{\rho N}^{\ast 2}}{4 m^{\ast 2}_{\rho}} \rho_3 \delta_{\mu, 0} \ , \label{eqn:S0}
\end{equation}
where the isoscalar density, $\rho_{\omega}$, is given by
\begin{equation}
\rho_{\omega}= \rho_p +\rho_n + \displaystyle \sum_H r^{\omega}_{HN} \rho_H  \label{eqn:vecd},
\end{equation} 
and $r^{\omega}_{HN} = g_{\omega H}/g^{\ast}_{\omega N}$ is the density-dependent ratio of hyperon-nucleon coupling
constants.   The self-energies,
$\Sigma^{\mu}_{\rho p}$ and $\Sigma^{\mu}_{\rho n}$, are briefly denoted as $\Sigma^{\mu}_{\rho(^p_n)}$; the isovector
density is denoted as $\rho_3 = (k^3_{F_p} - k^3_{F_n})/3\pi^2$ where the Fermi momentum $k_{F_p}$ is for proton and
$k_{F_n}$ for neutron~${\cite{UEC2,SCH}}$.  The baryon-isovector density, $\rho_{3B}$, and the ratios of sigma-nucleon coupling
constants on $\rho$-meson are also defined; for example, $\rho_{3B} = \rho_3 + r^{\rho}_{\Sigma N} \rho_{3\Sigma}$, where
$r^{\rho}_{\Sigma N}=g_{\rho \Sigma}/g_{\rho N}^{\ast}$ and $\rho_{3 \Sigma} = \rho_{\Sigma^{+}} - \rho_{\Sigma^{-}}$.

Thermodynamically consistent effective masses of mesons compatible with
effective coupling constants~($\ref{eqn:EFC}$) are required to be:
\begin{equation}
\begin{split}
m_{\sigma}^{\ast 2} =&\ m_{\sigma}^2 \Bigl( 1 + \frac{g_{\sigma 3}}{2 m_{\sigma}^2} \phi_0
+ \frac{g_{\sigma 4}}{3! m_{\sigma}^2} \phi_0^2 - \frac{g_{\sigma \omega}}{2 m_{\sigma}^2} V_0^2
- \frac{g_{\sigma \rho}}{2 m_{\sigma}^2} R_0^2 - \frac{g_{\sigma \sigma N}}{2m_{\sigma}^3} \rho_{sN} \Bigr) \ , \\
m_{\omega}^{\ast 2} =&\ m_{\omega}^2 \Bigl( 1 + \frac{g_{\omega 4}}{3! m_{\omega}^2} V_0^2
+ \frac{g_{\sigma \omega}}{2 m_{\omega}^2} \phi_0^2 + \frac{g_{\omega \rho}}{2 m_{\omega}^2} R_0^2 \Bigr) \ , \\
m_{\rho}^{\ast 2} =&\ m_{\rho}^2 \Bigl( 1 + \frac{g_{\rho 4}}{3! m_{\rho}^2} R_0^2
+ \frac{g_{\sigma \rho}}{2 m_{\rho}^2} \phi_0^2 + \frac{g_{\omega \rho}}{2 m_{\rho}^2} V_0^2 \Bigr) \ . \label{eqn:MVC}
\end{split}
\end{equation}
Since effective masses of mesons and coupling constants depend on mean fields of mesons, they are density-dependent
through meson fields.  Note that the effective mass of $\sigma$-meson depends on the ($n, p$) scalar source of
nucleons, $\rho_{sN}$.  The modifications to equations of motion, propagators and self-energies produced by density-dependent
effective coupling constants and masses have to be carefully discussed.

The energy density, pressure of isospin-symmetric, asymmetric and charge-neutral hadronic matter are calculated by way of the energy-momentum tensor:
\begin{equation}
\begin{split}
{\cal E}_{\scriptscriptstyle NHA} =& \sum_B \frac{1}{\pi^2} \int^{k_{F_B}}_0\! dk k^2 E_B(k)
+ \frac{m_{\sigma}^2}{2} \phi_0^2 + \frac{g_{\sigma 3}}{3!} \phi_0^3 + \frac{g_{\sigma 4}}{4!} \phi_0^4 
- \frac{m_{\omega}^2}{2} V_0^2 - \frac{g_{\omega 4}}{4!} V_0^4 - \frac{g_{\sigma \omega}}{4} \phi_0^2 V_0^2 \\
& - \Bigl( \frac{m_{\rho}^2}{2} + \frac{g_{\rho 4}}{4!} R_0^2 + \frac{g_{\sigma \rho}}{4} \phi_0^2 +
\frac{g_{\omega \rho}}{4} V_0^2 \Bigr) R_0^2 + \sum_{l = e^{-}, \mu^{-}} \frac{1}{\pi^2} \int^{k_{F_l}}_0\! dk k^2 E_l(k)
\ , \label{eqn:Epne}
\end{split}
\end{equation}
\begin{equation}
\begin{split}
p_{\scriptscriptstyle NHA} =& \frac{1}{3\pi^2} \sum_B \int^{k_{F_B}}_0\! dk \frac{k^4}{E^{\ast}_B (k)}
- \frac{m_{\sigma}^2}{2} \phi_0^2 - \frac{g_{\sigma 3}}{3!} \phi_0^3 - \frac{g_{\sigma 4}}{4!} \phi_0^4
+ \frac{m_{\omega}^2}{2} V_0^2 + \frac{g_{\omega 4}}{4!} V_0^4 + \frac{g_{\sigma \omega}}{4} \phi_0^2 V_0^2 \\
& + \Bigl( \frac{m_{\rho}^2}{2} + \frac{g_{\rho 4}}{4!} R_0^2 + \frac{g_{\sigma \rho}}{4} \phi_0^2 +
\frac{g_{\omega \rho}}{4} V_0^2 \Bigr) R_0^2 + \sum_{l = e^{-}, \mu^{-}} \frac{1}{3\pi^2} \int^{k_{F_l}}_0\! dk
\frac{k^4}{E_l^{\ast} (k)}\ , \label{eqn:Ppne}
\end{split}
\end{equation}
where $k_{F_B}$ is the Fermi-momentum; $E_B (k)$ and $E_l (k)$ are single particle energies for baryons and leptons,
respectively.  One can check that the thermodynamic
relations, such as ${\cal E}_{\scriptscriptstyle NHA} + p_{\scriptscriptstyle NHA} = \rhoB E_n (k_{F_n})$ and the chemical potential,
$\mu = \partial {\cal E}_{\scriptscriptstyle NHA}/ \partial \rhoB = E_n (k_{F_n}) = E^{\ast} (k_{F_n}) - \Sigma^0 (k_{F_n})$, are
exactly satisfied with a given baryon density, $\rhoB=2\kfermi^3/3\pi^2$.

The functional derivative of energy density, ${\cal E}_{\scriptscriptstyle NHA}(\phi_0, V_0, R_0, n_i)$, with respect to the baryon
number distribution, $n_i$, is given by:
\begin{equation}
\frac{\delta {\cal E}_{\scriptscriptstyle NHA}}{\delta n_i} = E(k_i) + \sum_i \Bigl( \frac{\delta {\cal E}_{\scriptscriptstyle NHA}}{\delta \phi_0}
\frac{\delta \phi_0}{\delta n_i} + \frac{\delta {\cal E}_{\scriptscriptstyle NHA}}{\delta V_0}
\frac{\delta V_0}{\delta n_i} + \frac{\delta {\cal E}_{\scriptscriptstyle NHA}}{\delta R_0} \frac{\delta R_0}{\delta n_i} \Bigr) \ . \label{eqn:dedn}
\end{equation}
Thermodynamic consistency requires: $\displaystyle \frac{\delta {\cal E}_{\scriptscriptstyle NHA}}{\delta \phi_0} = 0$,
$\displaystyle \frac{\delta {\cal E}_{\scriptscriptstyle NHA}}{\delta V_0} = 0$ and
$\displaystyle \frac{\delta {\cal E}_{\scriptscriptstyle NHA}}{\delta R_0} = 0$~$\cite{UEC1}$.  Now, one can directly prove
that the self-energies calculated by propagators and the conditions of conserving approximations become equivalent,
only if the effective masses and effective coupling constants of mesons are given
by~($\ref{eqn:EFC}$) and ($\ref{eqn:MVC}$)~$\cite{UEC2}-\cite{HUS}$.

In order to start self-consistent hadronic matter calculations, nonlinear coupling constants,
$g_{\sigma N}$, $g_{\omega N}$, $g_{\rho N}$ and other 9 nonlinear coefficients
should be supplied; but admissible values of nonlinear coefficients are determined to satisfy properties of symmetric nuclear matter
at saturation ($-15.75$ MeV at $\kfermi =1.30$ fm$^{-1}$, $a_4 = 30.0$ MeV), simultaneously searching for the minimum value of
incompressibility and reproducing the maximum mass of isospin-asymmetric neutron stars
($M_{\rm max} (n,p,e) = 2.50$ $M_{\odot}$).  In addition, hadronic phase transitions from ($n, p, e$) to
($n, p, H_1, H_2, \cdots, e$) should be carefully incorporated in EOS.  The binding energies of hyperons required by hadronic model
and SU(6) quark model for vector coupling constants~$\cite{JSI,GSY}$ are compared and discussed quantitatively
in the articles~$\cite{SCH,HUS}$.

\section{The nonlinear $\sigma$-$\omega$-$\rho$ mean-field approximation with vacuum fluctuation correction (VFC)}

We have included vacuum fluctuation correction (VFC) into the nonlinear $\sigma$-$\omega$-$\rho$ mean-field approximation and
applied the EOS with or without VFC to hadronic stars and hadron-quark hybrid stars.  The EOSs for hadrons with VFC and quark
matter generated by MIT-bag model are briefly discussed; based on the formalism, the pure hadronic and quark stars,
stability of hadron-quark stars, are discussed in sec.~6.  The vacuum fluctuation corrections are explicitly performed with
counterterms required by power-counting and the method of dimensional regularization~$\cite{WAL}$.

The self-energies with VFC in the conserving nonlinear $\sigma$-$\omega$-$\rho$ approximation are given with $\eqref{eqn:TSS}$ by,
\begin{equation}
\begin{split}
\Sigma^s_B =& M^{\ast}_B(\kfermi) - M_B \\
=& \Sigma^s_N + \Sigma^s_H + \frac{1}{2\pi^{2}m^{\ast 2}_{\sigma}} \sum_{B} g_{\sigma B}^{\ast 2}
\Big[ M_{B}^{\ast 3}\ln \left( \frac{M_{B}^{\ast }}{M_{B}} \right)
-M_B^2(M_{B}^{\ast}-M_{B})-\frac{5}{2}M_{B}(M_{B}^{\ast}-M_{B})^{2} \\
&- \frac{11}{6}(M_{B}^{\ast}-M_{B})^{3}\Big] \ , \label{eqn:SVFC}
\end{split}
\end{equation}
where $B=n,p,\Lambda,\Sigma^{-},\cdots$, and $g_{\sigma B}^{\ast 2} \equiv g_{\sigma B}^2$ for $B=\Lambda,\Sigma^{-},\cdots$,
since we are investigating effects of density-dependent interactions of nucleons; the density-dependent nonlinear self and
mixing interactions among hyperons will be studied in the future.  The self-energies $\Sigma^v$ and $\Sigma^0$ obtain no
VFC in the mean-field approximation, which will be proven directly by dimensional regularization method.  Note that it is also
essential for an approximation with VFC to maintain conditions of conserving approximations in order to obtain~\eqref{eqn:SVFC}.

The energy density with VFC is derived as,
\begin{equation}
\begin{split}
\Delta {\cal E}_{VFC}=&-\frac{1}{8\pi^{2}}\sum_{B}\Big[ M_{B}^{\ast 4}\ln \left( \frac{M_{B}^{\ast }}{M_{B}} \right)
+M_{B}^{3}(M_{B}-M_{B}^{\ast})-\frac{7}{2}M_{B}^2(M_{B}-M_{B}^{\ast })^{2}\\
&\hspace{2.0cm}+\frac{13}{3}M_{B}(M_{B}-M_{B}^{\ast })^{3}-\frac{25}{12}(M_{B}-M_{B}^{\ast })^{4}\Big] \ . \label{eqn:DVFC}
\end{split}
\end{equation}
Then, the total energy density is given with eq.~($\ref{eqn:Epne}$) as, 
\begin{equation}
{\cal E}_{VFC} = {\cal E}_{NHA} + \Delta {\cal E}_{VFC} \ . \label{eqn:EVFC}
\end{equation}
The self-energies and pressure are also evaluated by dimensional regularization and thermodynamic consistency can be proved
including VFC.  As discussed in sec.~2, all coupling constants have to be evaluated by maintaining properties at nuclear matter
saturation, searching the minimum value of incompressibility.  Incomressibility is $K \sim 350$ MeV and
$M_{max} = 2.33$ M$_{\odot}$ for ($n,p,e$)+VFC matter.  The vacuum fluctuation corrections to nonlinear $\sigma$-$\omega$-$\rho$
mean-field approximation are not significant in low and high densities compared with those of nonlinear
$\sigma$-$\omega$-$\rho$ self and mixing interactions.

The energy density and pressure of MIT-bag model are derived~$\cite{BDS}-\cite{EFA}$:
\begin{equation}
{\cal E}_q = \frac{3}{8\pi^2} \sum_f \Bigl[ 2k_f E_f^3(k_f) - m_f^2k_fE_f(k_f) - m_f^4 \log|\frac{k_f + E_f(k_f)}{m_f}| \Bigr ]
 + B \ , \label{eqn:Eqk}
\end{equation}
\begin{equation}
P_q = \frac{1}{8\pi^2} \sum_f \Bigl[ (2k_f^3-3k_fm_f^2) E_f(k_f) + 3m_f^4 \log|\frac{k_f + E_f(k_f)}{m_f}| \Bigr]
 - B \ , \label{eqn:Pqk}
\end{equation}
where $E_f(k_f) = (k_f^2 + m_f^2)^{1/2}$, ($f = u, d, s$), and $B$ is the bag-constant (MeV/fm$^3$); $k_f$ is the Fermi-momentum
of flavor $f$.  The energy density and pressure with the following given baryon density, charge neutrality and
phase equilibrium conditions:
\begin{equation}
\rhoB = \frac{2\kfermi^3}{3\pi^2} = \frac{1}{3} (\rho_u + \rho_d + \rho_s) \ , \label{eqn:BNC}
\end{equation}
\begin{equation}
0= 2\rho_u - \rho_d - \rho_s  \ , \label{eqn:CN}
\end{equation}
\begin{equation}
E_d (k_d) = E_s(k_s) \ , \label{eqn:QPE}
\end{equation}
determine ($u,d,s$)-quark matter uniquely (note: $\rho_f = k_f^3/\pi^2$).  The thermodynamic potential for $f$ to first order
in the strong coupling constant is given by~$\cite{ACH,EFA,GLE}$:
\begin{equation}
\begin{split}
{\Omega_f} &= -\frac{\gamma_f}{24\pi^2} \biggl\{ k_f E_f(k_f)(k_f^2-\frac{3}{2}m_f^2) + \frac{3}{2} m_f^4 \log|\frac{k_f + E_f(k_f)}{m_f}| \\
&-\frac{2\alpha_s}{\pi} \Bigl[ 3\Bigl(E_f(k_f)k_f - m_f^2 \log|\frac{k_f + E_f(k_f)}{E_f(k_f)}| \Bigr)^2 -2k_f^4 -3m_f^4\log^2
\Bigl( \frac{m_f}{E_f(k_f)} \Bigr) \\
&+ 6\log \Bigl( \frac{\sigma}{E_f(k_f)} \Bigr) \Bigl( k_fE_f(k_f)m_f^2 - m_f^4 \log|\frac{k_f + E_f(k_f)}{m_f}| \Bigr) \Bigr]
\biggr\} \ , \label{eqn:FOC}
\end{split}
\end{equation}
where $\gamma_f = spin \times color$ degeneracy.  $\alpha_s$ is the strong interaction coupling constant, and $\sigma$ is the renormalization
scale constant which is considered as a typical chemical potential of the problem and chosen as $\sim 300$ MeV~$\cite{GLE}$.  However,
the EOS of hadron-quark compact stars is not sensitive to the parameter $\sigma$, which is numerically checked and theoretically expected
from $\eqref{eqn:FOC}$ in high density region.  It is suggested that the EOS of quark matter interconnected
with hadronic matter be mainly sensitive to parameters, ($B, \alpha_c$) in the current analysis.  The expected value and restriction
to ($B, \alpha_c$) can be numerically extracted from properties of hadron-quark matter and compact stars in the sec.~6.

The energy density and pressure are now expressed as,
\begin{equation}
\begin{split}
{\cal E}_q &= B + \sum_f ( \Omega_f + \mu_f n_f) \ ,\\
P_q &= -B - \sum_f \Omega_f  \ , \label{eqn:QEP}
\end{split}
\end{equation}
where $\mu_f = E_f(k_f)$ and $n_f$ are chemical potential and particle distribution for the flavor $f$.  The equations of state for hadronic
and quark matter, phase transition conditions and TOV equation~$\cite{WEI}$ are employed to calculate pure-hadron, pure-quark and
hadron-quark compact stars.

\section{The properties of hadronic and hadron-quark (H-Q) hybrid stars}

We have applied the conserving nonlinear $\sigma$-$\omega$-$\rho$ mean-field approximation with a VFC to examine
properties of hadronic neutron stars and H-Q hybrid stars.

The properties of pure hadronic stars are produced by employing hadronic equations of state, phase transition conditions,
self-consistent effective masses and coupling constants, and the TOV equation.  The masses v.s. central densities of hadronic stars
produced by EOSs for ($n, p, e$)-($n, p, \Sigma^{-}, e$)-($n, p, \Sigma^{-}, \Lambda$) matter are shown in
Figs.~1a ($r^{\omega}_{\Lambda N}=1.0, r^{\omega}_{\Sigma N}=1.0$) and
1b ($r^{\omega}_{\Lambda N}=2/3, r^{\omega}_{\Sigma N}=2/3$).   The solid line shows ($n, p, e$) matter which produces the
maximum mass of hadronic neutron stars, $M_{max} = 2.50$ M$_{\odot}$~$\cite{SCH}$.   LHA(n) (dot-dashed line) is the result of the
linear $\sigma$-$\omega$ mean-field approximation~$\cite{WAL}$ ($M_{max} = 3.06$ M$_{\odot}$).  The arrows respectively indicate
phase transitions from ($n, p, e$)-($n, p, \Sigma^{-}, e$) and ($n, p, \Sigma^{-}, e$)-($n, p, \Sigma^{-}, \Lambda$) matter.  The
EOS of hadronic matter with $r^{\omega}_{\Lambda N}=1.0, r^{\omega}_{\Sigma N}=1.0$ generates the maximum mass
$M_{max}(n,p,\Sigma^{-},\Lambda) = 1.98$ M$_{\odot}$ (Fig.~1a).  However, the EOS with
$r^{\omega}_{\Lambda N}=2/3, r^{\omega}_{\Sigma N}=2/3$ as suggested by the SU(6) quark model, produces the small maximum mass
of neutron stars, $M_{max}(n,p,\Sigma^{-},\Lambda) = 1.33$ M$_{\odot}$ (Fig.~1b).   This is unable to support the observed masses 
of neutron stars.  The discrepancy calculated by $r^{\omega}_{H N}=1.0$ and $r^{\omega}_{H N}=2/3$ is clearly recognized
by comparing the EOS and mass of neutron stars.

\begin{figure}[t]
\noindent\begin{minipage}[t]{0.50\textwidth}
\begin{center}
\includegraphics[height=.25\textheight]{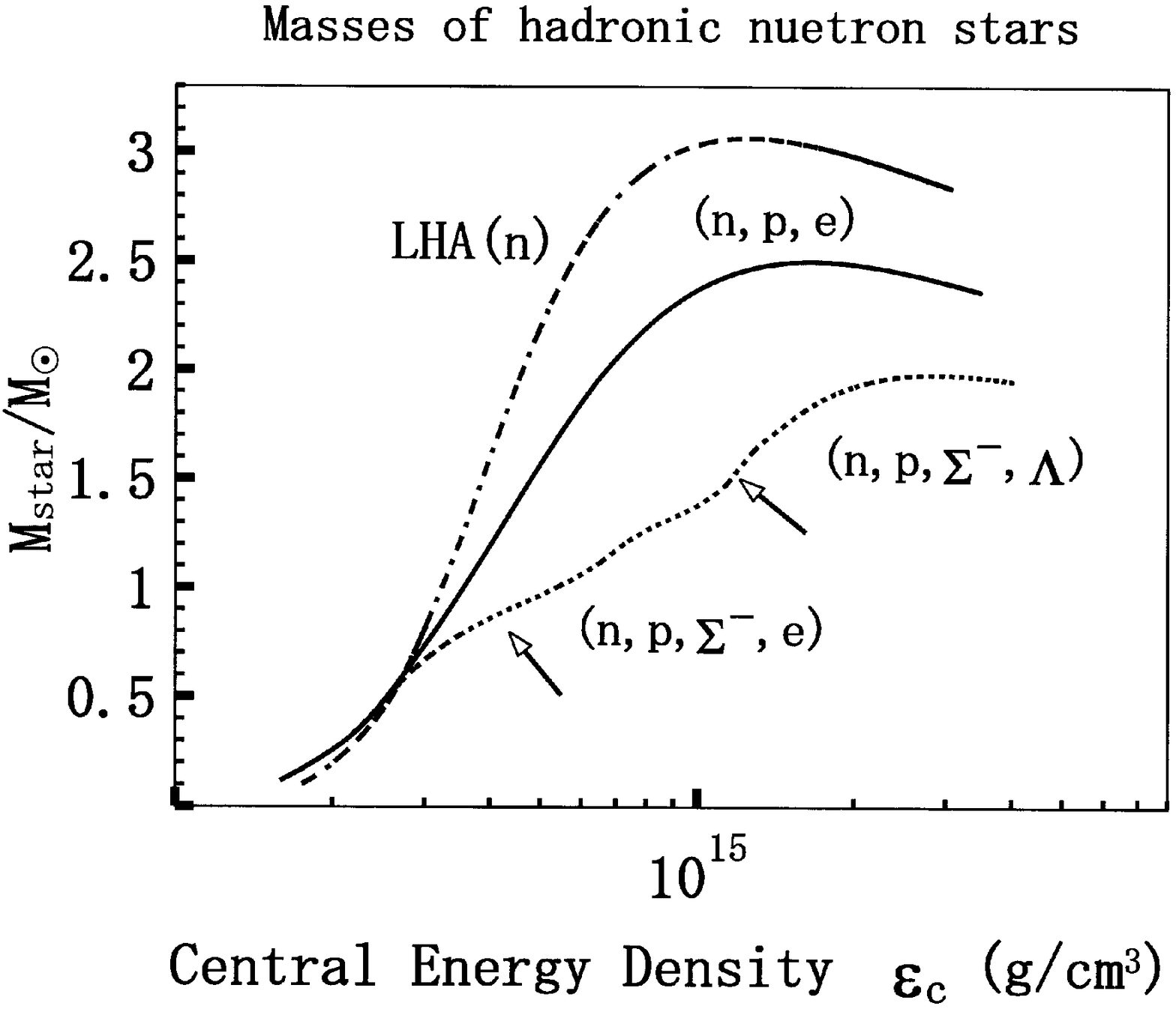}
\label{fig9a}
\end{center}
{Fig.~1a. Masses of hadronic neutron stars.  Pure-neutron matter in linear $\sigma$-$\omega$:
$M(n)_{max} = 3.06$ $M_{\odot}$.  Isospin asymmetric $\beta$-equilibrium matter: $M(n,p,e)_{max} = 2.50$
$M_{\odot}$.  ($n, p, e$)-($n, p, \Sigma^{-}, e$)-($n, p, \Sigma^{-}, \Lambda$) matter with the coupling
ratios: ($r^{\sigma}_{\Lambda N}=0.964$, $r^{\omega}_{\Lambda N}=1.00$) and
($r^{\sigma}_{\Sigma^{-} N}=0.925$, $r^{\omega}_{\Sigma^{-} N}=1.00$, $r^{\rho}_{\Sigma^{-} N}=1.00$)
give $M_{max} = 1.98$ $M_{\odot}$.}
\end{minipage}\quad
\begin{minipage}[t]{0.50\textwidth}
\begin{center}
\includegraphics[height=.25\textheight]{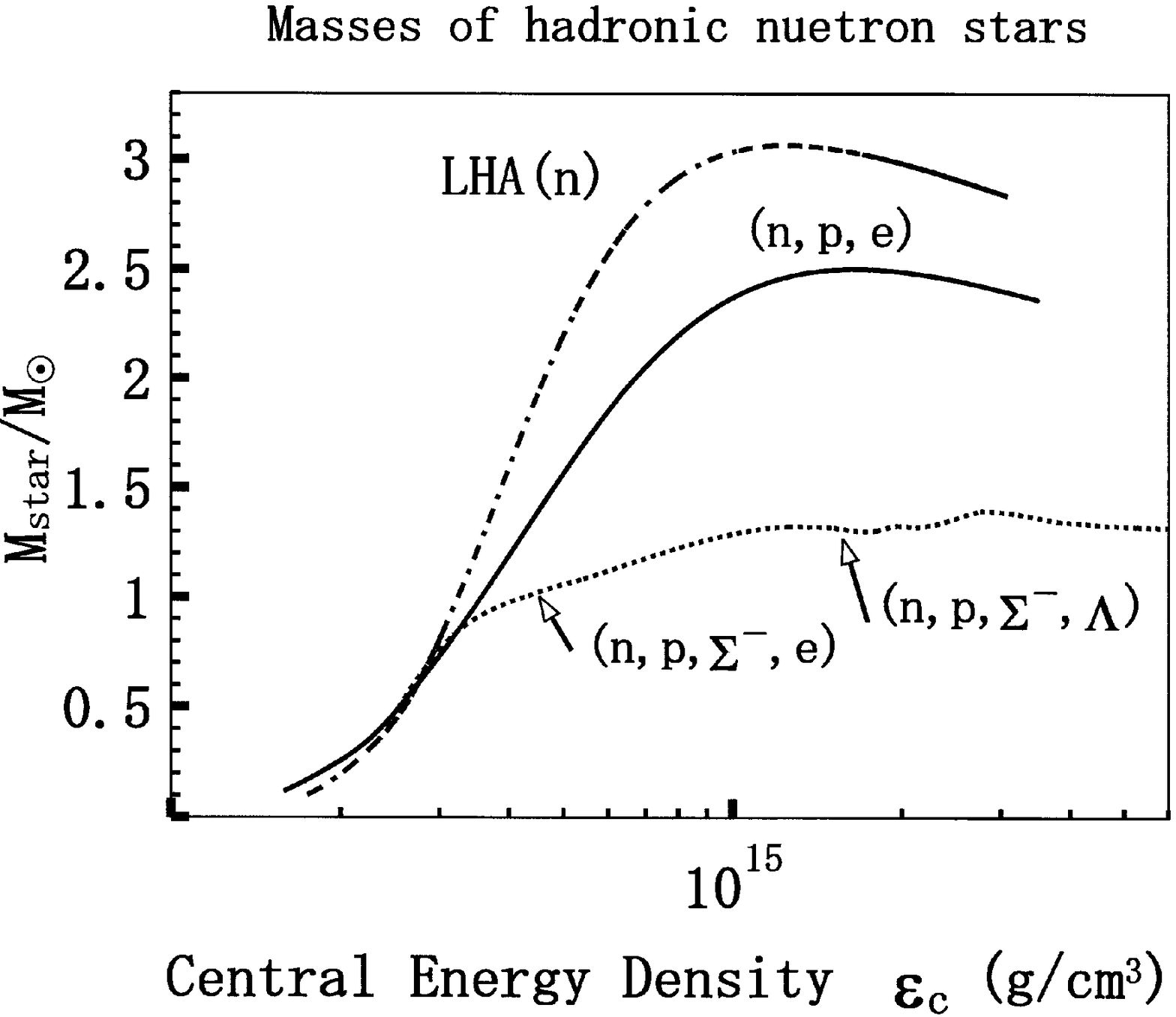}
\label{fig9b}
\end{center}
{Fig.~1b.  Masses of hadronic neutron stars.  ($n, p, e$)-($n, p, \Sigma^{-}, e$)-($n, p, \Sigma^{-}, \Lambda$)
matter with the coupling ratios: ($r^{\sigma}_{\Lambda N}=0.677$, $r^{\omega}_{\Lambda N}=2/3$)
and ($r^{\sigma}_{\Sigma^{-} N}=0.632$, $r^{\omega}_{\Sigma^{-} N}=2/3$, $r^{\rho}_{\Sigma^{-} N}=1.00$)
give $M_{max} = 1.40$ $M_{\odot}$.}
\end{minipage}
\end{figure}

All curves of ${\cal E}_c - M_{\rm star}$ in Fig.~1a and 1b have simple positive slopes, $dM/d{\cal E}_c > 0$, and one maximum
point which is typical for lines of one critical (inflection) point, $d^2M/d{\cal E}_c^2 < 0$.  In this case, stability can be
examined clearly for these simple saturating curves by the condition $dM_{star}/d{\cal E}_c > 0$~$\cite{GLE,SLS}$,
respectively.  Hence, values of positive slopes of ${\cal E}_c - M_{\rm star}$ are solutions
of stable neutron stars, but the decreasing curve, $dM_{star}/d{\cal E}_c > 0$, after the maximum point indicates unstable
neutron stars.  These characters of ${\cal E}_c - M_{\rm star}$ are important to examine stability of hadron-quark stars.

The EOS of pure-quark matter is used to calculate masses of pure-quark stars, as shown in Fig.~2.  The bag constant
should be $B \gtrsim 80$ MeV/fm$^3$, as normal nuclear matter becomes quark matter in the case of
$B \simeq 80$ MeV/fm$^3$.  The maximum masses of quark stars are $M_{max} = 1.59$ M$_{\odot}$ ($B = 80$ MeV/fm$^3$),
$M_{max} = 1.31$ M$_{\odot}$ ($B = 120$ MeV/fm$^3$) and $M_{max} = 1.18$ M$_{\odot}$ ($B = 150$ MeV/fm$^3$).  The
${\cal E}_c - M_{\rm star}$ curves in Fig.~2 and~3 for the current pure hadron and quark stars are smoothly increasing
and saturating.   These curves are classified as those of one critical (infrection) point.  Therefore, stability
of neutron stars in pure-hadron and pure-quark matter can be checked by the condition: $dM_{\rm star}/d{\cal E}_c > 0$, though
one needs to check other conditions if ${\cal E}_c - M_{\rm star}$ curves exhibit  complicated phase transitions so that
critical points are larger than 2~$\cite{SLS}$.

\begin{figure}[t]
\begin{center}
\includegraphics[height=.25\textheight]{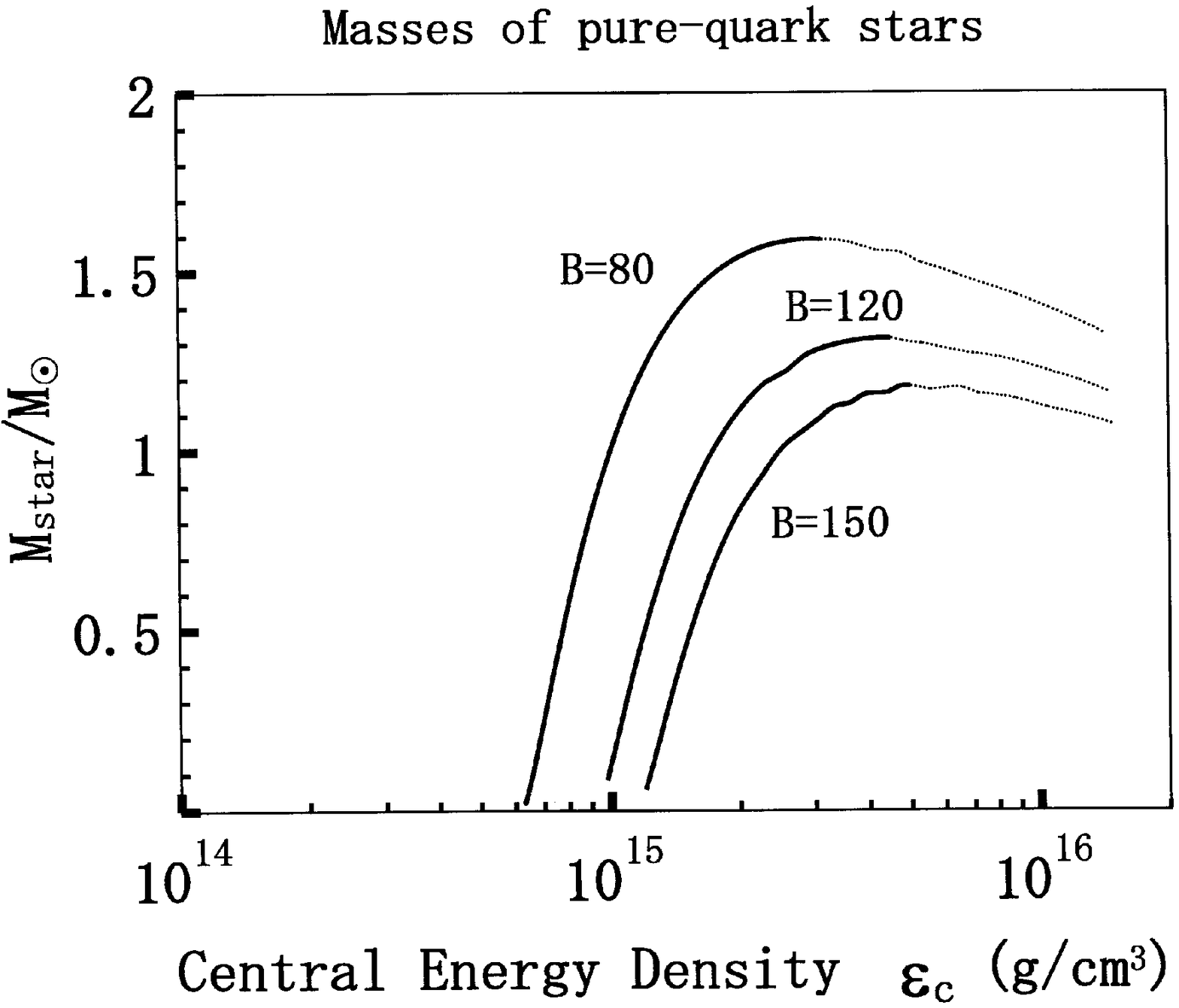}
\label{fig11}
\end{center}
{Fig.~2. Masses of pure-quark stars with $B= 80, 120, 150$ MeV/fm$^3$, ($\alpha_c = 0$).  The stable pure-quark stars
with $M_{max} \gtrsim 1.0$ are in $10^{15} \sim {\cal E}_c \sim 5.0 \times 10^{15}$ g/cm$^3$.}
\end{figure}

\begin{figure}[t]
\begin{center}
\includegraphics[height=.25\textheight]{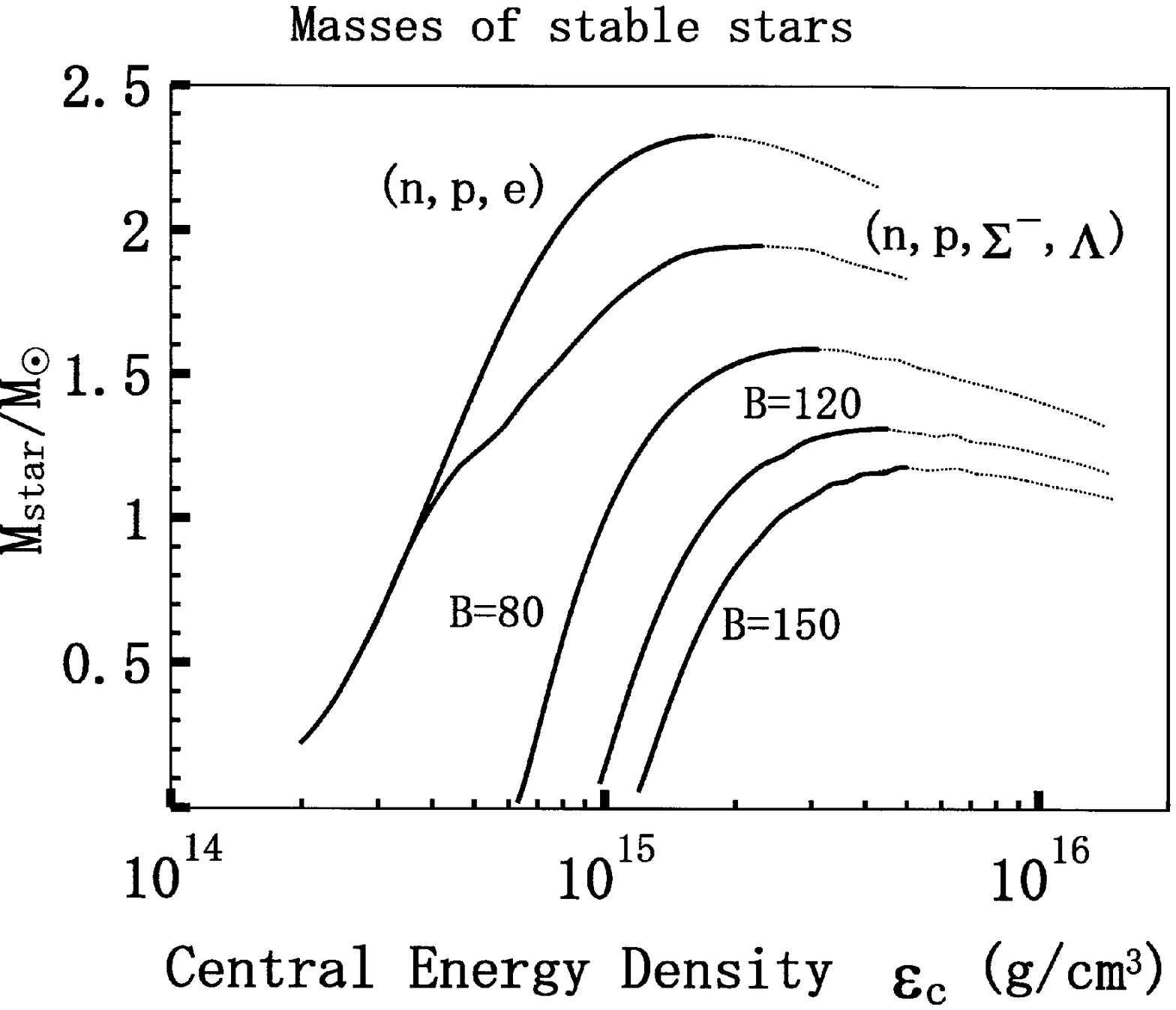}
\label{fig12}
\end{center}
{Fig.~3. Hadron stars with VFC and pure-quark stars.  The stable hadronic stars are
in $5.0 \times 10^{14} \sim {\cal E}_c \sim 10^{15}$ g/cm$^3$.}
\end{figure}

Stable quark stars with $M_{star} \gtrsim 1.0$ M$_{\odot}$ are limited in the range of central energy densities:
$10^{15} \lesssim {\cal E}_c \lesssim 5.0 \times 10^{15}$ g/cm$^3$, as shown in bold, solid lines in Fig.~2.  Stable hadronic
neutron stars with the VFC and pure quark stars are shown in Fig.~4, to remark on the energy-density regions of stable
stars.  The results of hadronic and quark stars in Fig.~3 suggest that H-Q hybrid stars are possible about
the central energy density, ${\cal E}_c \sim 10^{15}$ g/cm$^3$, as the EOS of quark matter could be energetically preferable
at high densities, compared with that of hadrons.  However, the overlap in central energy density between stable
hadronic stars ($M^H_{max} \gtrsim 1.0$) and pure-quark stars ($M^Q_{max} \gtrsim 1.0$) is very narrow.  If stable central energy
densities are the same during the H-Q phase transition, this indicates that H-Q stars would become immediately unstable~$\cite{BDS}$.

We have assumed that the H-Q phase transition is a first-order, and solved phase transition conditions for chemical potential
and pressure ($\mu_H = \mu_Q$ and $P_H = P_Q$) by employing the double-tangent method numerically.  The TOV equation,
as well as hadronic and quark EOSs with phase transition conditions, are applied to calculate the properties
of neutron stars.   First, we have examined H-Q hybrid stars in case of ($n, p, e$) + VFC and quark matter.  The results
are shown in Fig.~4.  In this figure, the mass of H-Q stars is shown with a solid line and the stable quark core is
indicated by a dotted line.  The results are not sensitive to the parameter $\sigma$, which is checked by changing
$\sigma = 200 \sim 800$ MeV.  However, the H-Q stars are sensitive to the values of $B$ and $\alpha_c$,
$B = 100 \sim 150$ MeV/fm$^3$, $\alpha_c = 0.2 \sim 0.1$ to produce the observed masses of neutron stars
($M_{star} \gtrsim 1.30$ M$_{\odot}$).   This is because the quark-EOS shifts to high energy densities
if $B$ is increased, meaning that the EOS becomes softer in terms of pressure, resulting in compact stars with a smaller mass and radius
at high densities.  In addition, if the QCD coupling constant $\alpha_c$ is increased as $\alpha_c = 0.1 \rightarrow 0.2$,
the masses of stable H-Q stars become small and shifted to high densities.  Hence, appropriate values of $B$ and $\alpha_c$ to produce
the observed data of neutron stars in the current EOS are correlated to each other so that if $B$ is increased, $\alpha_c$ should be
decreased, such as in ($B \sim 100$ MeV/fm$^3$, $\alpha_c \sim 0.2$) and ($B \sim 150$ MeV/fm$^3$, $\alpha_c \sim 0.1$).  The
property of ($B, \alpha_c$) in the analysis of H-Q infinite matter agrees with the results of bag-model fits to
light-hadron spectra and renormalization group analyses in the paper by Farhi and Jaffe~$\cite{EFA}$.

\begin{figure}[t]
\begin{center}
\includegraphics[height=.25\textheight]{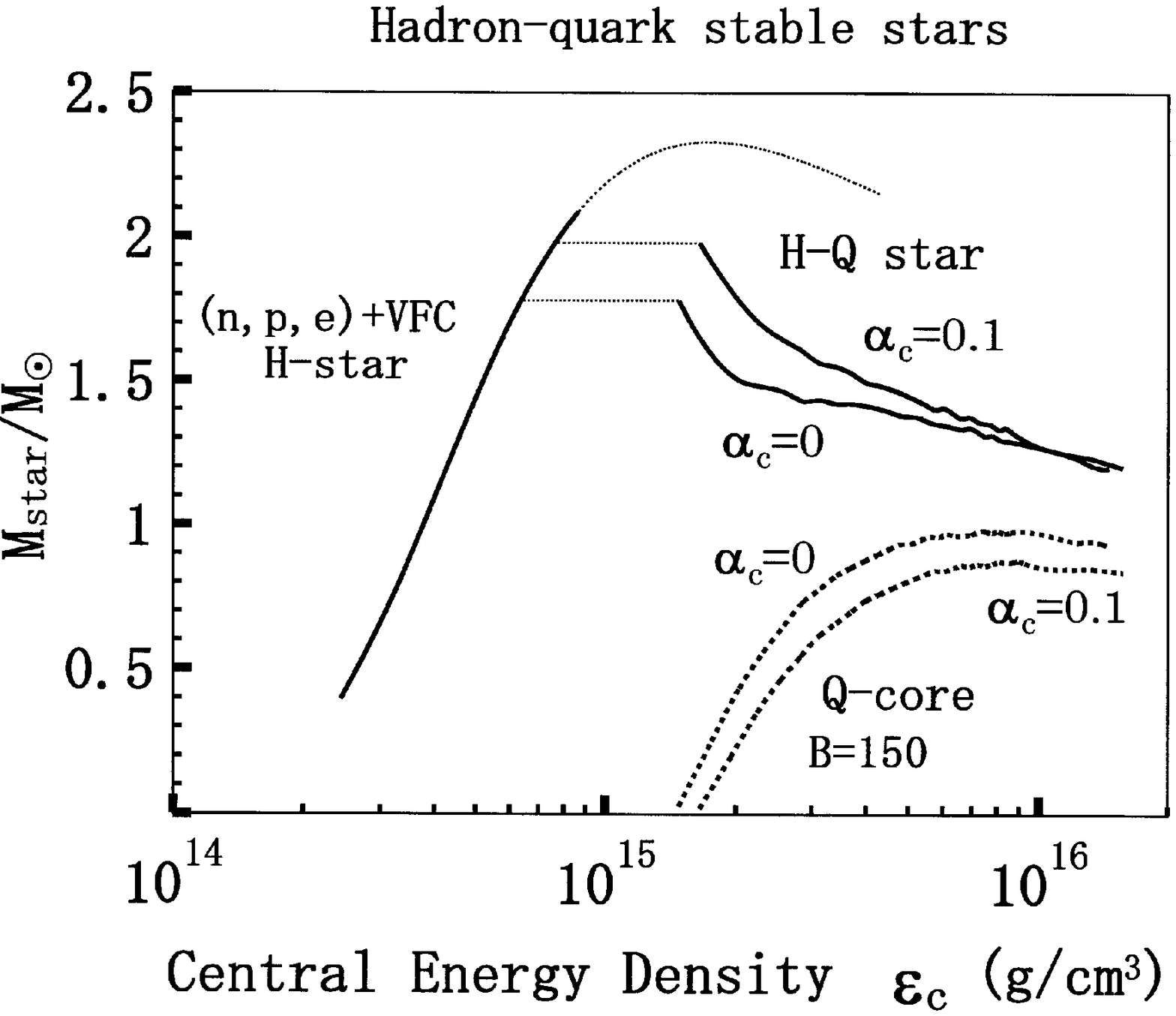}
\label{fig13}
\end{center}
{Fig.~4. Hadron-quark stars based on isospin asymmetric ($n, p, e$)+VFC matter.  Note that the quark-core of H-Q stars
is stable.}
\end{figure}

The stability of pure hadronic and quark matter is examined, respectively, by the condition
$dM_{star}/d{\cal E}_c > 0$~$\cite{GLE,SLS}$.  One should note that the stability criterion, $dM_{star}/d{\cal E}_c > 0$,
is for single phase compact stars, such as pure hadronic and quark stars.  As H-Q stars are 2-phase compact stars ({\it i.e.},
the quark phase for a star's core and hadron phase for a mantle), the stability of H-Q stars should be reconsidered.  

The H-Q stars in Fig.~4 show that the stable hadronic stars ($dM_H/d{\cal E}_c > 0$) in the central
energy range, ${\cal E}_c = 10^{14} \sim 10^{15}$ g/cm$^3$, will undergo a phase transition (dotted horizontal line), reaching
H-Q stars ($B = 150$ MeV/fm$^3$).  The Fig.~4 indicates that the total mass of H-Q stars decreases, but a stable quark-core
($dM_Q/d{\cal E}_c > 0$) develops.  Therefore, this suggests that the H-Q star is stable, although the total mass of the star
becomes smaller.  Moreover, by comparing stable energy densities of the quark phase in Fig.~3 with those of H-Q stars,
the central energy density
of stable H-Q stars is found to be more extended for higher densities than that of single phase stars.  This suggests that
compact stars consisting of a mantle and a high density core are more stable than stars in a homogeneous single phase
structure~$\cite{EFA,EDW}$.  When the QCD coupling constant, $\alpha_c$, increases, the H-Q phase transition density
and quark-core will shift to higher densities, but the quark-core is stable and extends to higher densities.  If the bag constant
is small, such as $B \sim 100$ MeV/fm$^3$, one will obtain a saturation curve (an inflection point for stability)
within $10^{15} \sim 10^{16}$ g/cm$^3$, as in Fig.~3.

The H-Q stars calculated with the EOS for ($n, p, e$)-($n, p, \Sigma^{-}, e$)-($n, p, \Sigma^{-}, \Lambda$) + VFC to
quark matter ($B = 150$ MeV/fm$^3$) are shown in Fig.~5a ($r^{\omega}_{H N}=1.0$) and 5b ($r^{\omega}_{H N}=2/3$),
respectively.  The maximum masses in Fig.~5a are $M_{max} = 1.61$ M$_{\odot}$ ($\alpha_c = 0.0$) and
$M_{max} = 1.77$ M$_{\odot}$ ($\alpha_c = 0.1$); whereas, in Fig.~5b, they are
$M_{max} = 1.37$ M$_{\odot}$ ($\alpha_c = 0.0$) and $M_{max} = 1.48$ M$_{\odot}$ ($\alpha_c = 0.1$).  A remarkable feature
of H-Q stars in Fig.~5a ($r^{\omega}_{H N}=1.0$) is that they can reasonably explain central energy densities and maximum
mass configurations~$\cite{LAT}$ for stable compact stars, ranging from $10^{14} \sim 10^{16}$ g/cm$^3$.

In the case of the softer hadronic EOS in Fig.~5b ($r^{\omega}_{H N}=2/3$)
with $B = 100 \sim 150$ and $\alpha_c$-correction, phase transitions occur in relatively
low densities.  This results in smaller maximum masses of H-Q stars, $M_{max} \lesssim 1.4$ M$_{\odot}$.  Thus,
the EOS with the coupling constants ($r^{\omega}_{H N}=2/3$) may not be appropriate to explain the observed masses of
neutron stars.   A physical reason from the hadronic sector is clear in terms of the EOS; however, it is very interesting to investigate
how hadronic and quark models reconcile the problems pointed out in the paper.

\begin{figure}[t]
\noindent\begin{minipage}[t]{0.50\textwidth}
\begin{center}
\includegraphics[height=.25\textheight]{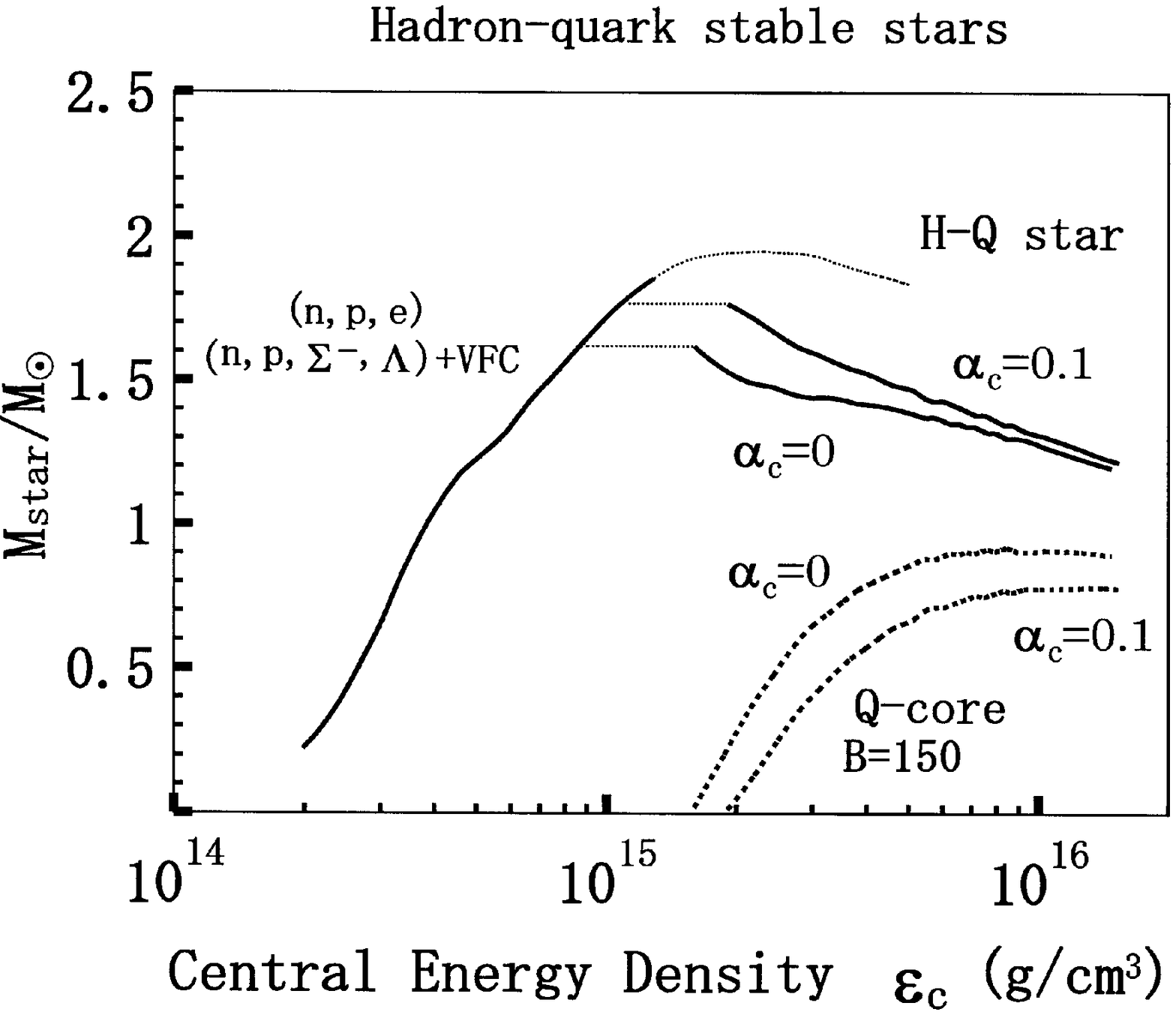}
\label{fig14a}
\end{center}
{Fig.~5a. Hadron-quark stars based on ($n, p, e$)-($n, p, \Sigma^{-}, e$)-($n, p, \Sigma^{-}, \Lambda$) + VFC
matter.  The coupling ratios: ($r^{\sigma}_{\Lambda N}=0.964$, $r^{\omega}_{\Lambda N}=1.00$) and
($r^{\sigma}_{\Sigma^{-} N}=0.925$, $r^{\omega}_{\Sigma^{-} N}=1.00$, $r^{\rho}_{\Sigma^{-} N}=1.00$).}
\end{minipage}\quad
\begin{minipage}[t]{0.50\textwidth}
\begin{center}
\includegraphics[height=.25\textheight]{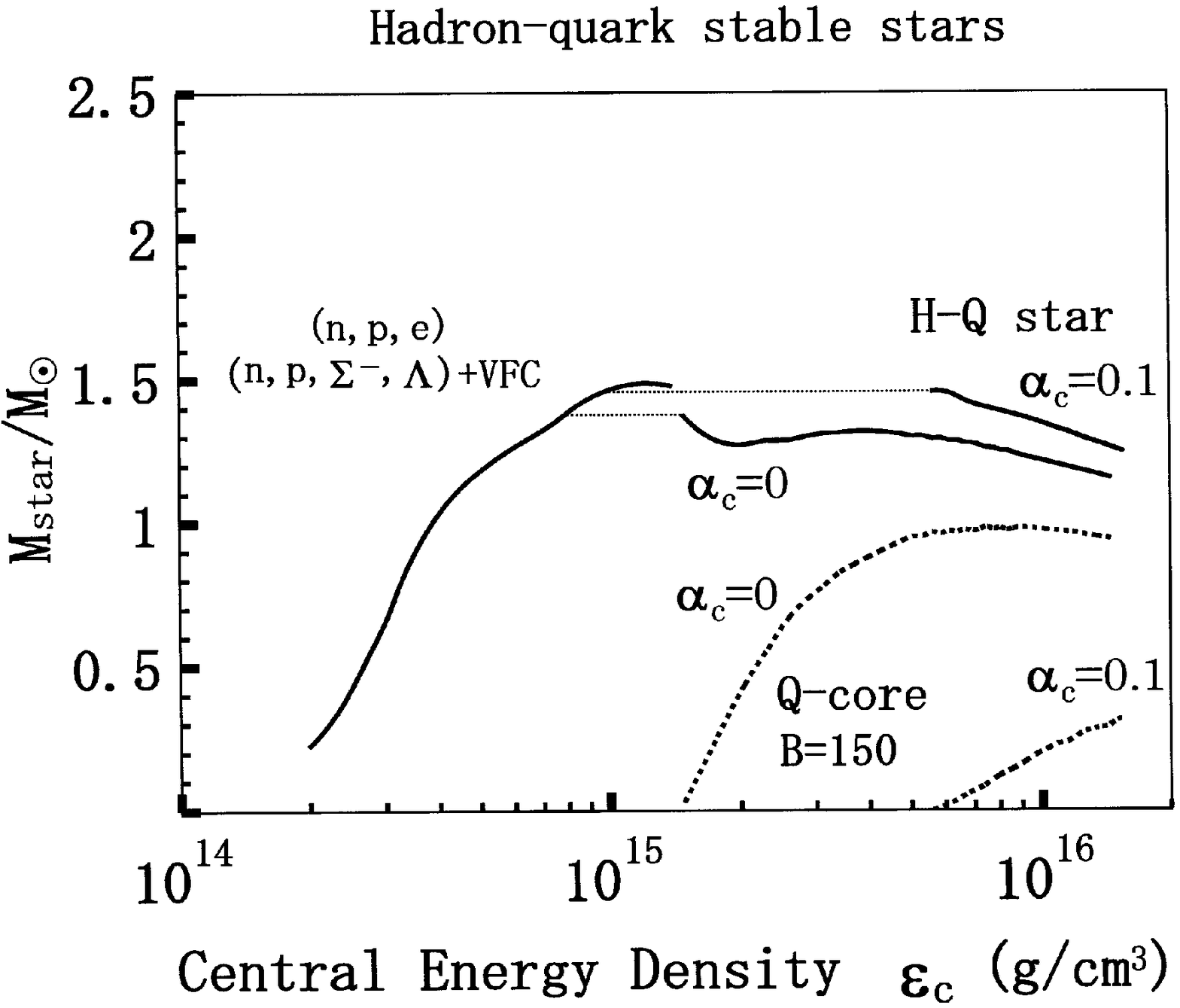}
\label{fig14b}
\end{center}
{Fig.~5b.  Hadron-quark stars based on ($n, p, e$)-($n, p, \Sigma^{-}, e$)-($n, p, \Sigma^{-}, \Lambda$) + VFC
matter.  The coupling ratios: ($r^{\sigma}_{\Lambda N}=0.677$, $r^{\omega}_{\Lambda N}=2/3$) and
($r^{\sigma}_{\Sigma^{-} N}=0.632$, $r^{\omega}_{\Sigma^{-} N}=2/3$, $r^{\rho}_{\Sigma^{-} N}=1.00$).}
\end{minipage}
\end{figure}

If the QCD coupling constant is set large ($\alpha_c \gtrsim 0.1$) in the soft EOS of Fig.~5b ($r^{\omega}_{H N}=2/3$), generation
of the quark-phase will move up to a very high density, separating the energy densities of hadron and quark phases completely.  In this case,
no stable hadron-phase can exist or coexist even in a crust of the surface of quark stars, resulting in literally
pure-quark stars.  Although more precise and detailed analyses are needed for many-body interactions of hadronic and
quark matter, the coupling constants for hadrons ($r^{\omega}_{H N}=2/3$ or $r^{\omega}_{H N}=1/3$), as suggested by the
SU(6) quark model, would not be appropriate to examine the properties of nuclear and hyperonic matter, or the maximum masses
of high density compact stars.

\section{Concluding remarks}

The quantum hadrodynamics (QHD) and mean-field approximations are interesting for describing hadronic many-body systems at
large distances, but they must ultimately break down at short distances where QCD is valid; hence, it is stimulating to investigate
how properties of hadron and quark dynamics will interconnect and influence each other dynamically~$\cite{WAL,SER}$.  We have applied
the conserving nonlinear $\sigma$-$\omega$-$\rho$ mean-field approximation to study self-consistent, density-dependent
interactions of hadrons and examined hadronic dynamics by employing coupling constants required from the QCD degrees of
freedom.  Discrepancies between hadronic and QCD predictions are shown in terms of effective masses, effective coupling constants,
incompressibility and symmetry energy, saturation properties of hyperon binding energies and maximum masses of high density
compact stars.  One of our purposes is to compare predictions indicated by effective hadronic and quark models so as to
clarify interrelations and distinctions between them.

As saturation properties of symmetric nuclear matter are self-consistently related to those of hyperons, reproducing
saturation properties of hyperons should be one of fundamental problems of nuclear many-body approximations~$\cite{DAY}-\cite{BAY}$.  Although
more quantitative and model-independent analyses are needed, it is concluded that
hyperon coupling ratios required by SU(6) quark model for the vector coupling constants~$\cite{JSI,GSY}$ are not appropriate
in order to generate saturation properties of hyperons.  The coupling ratios of hyperons, $r^{\omega}_{H N} \sim 1.0$, are appropriate
to explain Fermi-liquid properties, saturation of hyperons~$\cite{SCH,HUS}$ and maximum masses of neutron stars.  Discrepancies of
hyperon coupling ratios required by hadronic and SU(6) quark models should indicate that both effective approaches be improved
with each other as consistent theories for hadronic physics.

Heavy-ion collision experiments as well as neutron stars are useful to examine applications of quark models and
constraints for hadronic calculations~$\cite{VAD,IBO}$, since conditions of hadron-quark phase transition depend on both
equations of state for hadronic and quark matter.  The appropriate values of ($B, \alpha_c$) can be independently suggested
from properties of infinite matter, which will qualitatively support properties discussed in the paper~$\cite{EFA}$.  The stable
pure-hadronic stars and pure-quark stars based on MIT-bag model exist in different energy densities respectively, but
if a first-order phase transition is assumed, the energy density of stable hadron-quark stars expands from
$10^{14}$ to $10^{16}$ g/cm$^3$.  In the range of energy density, pure-hadron stars are less than $M(n,p,e)_{max} \lesssim 2.5$, but
H-Q stars are $M^{H-Q}_{max} \gtrsim 1.0$ $M_{\odot}$.  It reasonably explains expected masses and energy density relations,
$M-{\cal E}_c$~$\cite{LAT}$.  Therefore, existing high density compact stars are more likely to be regarded as hadron-quark
stars rather than pure-hadron and pure-quark stars.  These results should be further examined quantitatively from empirical data
and effective theories of hadrons.

It is noteworthy that density-dependent many-body effects simulated by nonlinear $\sigma$-$\omega$-$\rho$ interactions
are more important than those of VFC in the nonlinear $\sigma$-$\omega$-$\rho$ mean-field approximation.  The contributions
of VFC should be investigated further in more complicated nonlinear $\sigma$-$\omega$-$\rho$ HF, BHF approximations in order to study vacuum-fluctuation and density-dependent corrections.  The role of chiral symmetry in hadronic models should be examined by
extending the current nonlinear $\sigma$-$\omega$-$\rho$ mean-field approximation by including $\pi$-meson, which would reveal
physical meaning and significance of chirality for hadronic and quark models.


\end{document}